\documentclass[11pt,twoside
]{article}

\usepackage{baaa2008}
\usepackage{graphicx}
\usepackage{subfigure}
\usepackage{psfrag}
\usepackage{amssymb}
\usepackage[spanish,activeacute,english]{babel}
\usepackage[latin1]{inputenc}
\usepackage[T1]{fontenc} 
\usepackage{ae,aecompl} 
\usepackage{latexsym}
\usepackage{verbatim}
\usepackage{amsmath}
\usepackage{stmaryrd}
\usepackage{amsfonts}
\usepackage{amssymb}
\usepackage{wasysym}
\usepackage[colorlinks=true,dvips]{hyperref}

\begin{document}
\myselectenglish
\vskip 1.0cm
\markboth{A. V. Smith Castelli et al. }%
{A common colour-magnitude relation from ellipticals to globular clusters?}

\pagestyle{myheadings}
\vspace*{0.5cm}
\parindent 0pt{PRESENTACI\'ON MURAL}
\vskip 0.3cm
\title{A common colour-magnitude relation from giant elliptical galaxies to 
globular clusters? {}}

\author{A.V. Smith Castelli $^{1,2,3}$, L. P. Bassino $^{1,2,3}$, T. Richtler 
$^{4}$, F. Faifer $^{1,2,3}$, J. C. Forte $^{1,3}$, S. A. Cellone $^{1,2,3}$}

\affil{%
  (1) Facultad de Ciencias Astron\'omicas y Geof\'isicas - UNLP\\
  (2) IALP (CCT La Plata - CONICET - UNLP)\\
  (3) CONICET \\
  (4) Universidad de Concepci\'on - Chile\\ 
}

\begin{abstract} We discuss the existence of a common colour-magnitude 
relation (CMR) of metal-poor globular clusters and early-type 
galaxies, i.e. giant ellipticals, normal ellipticals and lenticulars, dwarf 
ellipticals and lenticulars, and dwarf spheroidals. Such CMR would cover a 
range of $\sim 14$ mag, extending from the brightest galaxies, down to the 
globular clusters on the fainter side. 
\end{abstract}

\begin{resumen}
Se investiga sobre la existencia de una \'unica relaci\'on color-magnitud  
trazada por c\'umulos globulares pobres en metales y por galaxias de tipo 
temprano, i.e. el\'ipticas gigantes, el\'ipticas y lenticulares normales, 
el\'ipticas y lenticulares enanas, y enanas esferoidales. Tal relaci\'on 
cubrir\'ia un rango de $\sim 14$ mag, y se extender\'ia desde las galaxias 
m\'as brillantes hasta los c\'umulos globulares en el extremo d\'ebil.
\end{resumen}

\section{Introduction}

The fact that early-type galaxies in clusters and groups define a tight 
sequence in the colour-magnitude diagram (CMD) in the sense that bright 
ellipticals are redder than fainter ones, has been known for a long time 
(e.g. Baum 1959). Spectroscopic studies of giant ellipticals (e.g. Kuntschner 
2000) and dwarf galaxies (e.g. Mieske et al. 2007) have shown that this 
colour-magnitude relation (CMR) mainly reflects metallicity effects: the more 
luminous (i.e. massive) galaxies are more capable to retain their metal 
content than low-mass ones thanks to their deeper potential wells (e.g. Rakos 
et al. 2001, and references therein). 

Studies of extragalactic globular cluster (GC) systems observed with HST have 
revealed that in some galaxies, the more luminous blue (i.e. metal-poor) GCs 
are redder than fainter ones, showing what seems to be a CMR. This relation 
has been called {\it blue tilt} and its origin and existence are still 
under discussion (Strader \& Smith 2008, and references therein).

\section{The ``blue tilt'' of globular clusters in NGC\,4486}

From ACS-HST observations, the existence of a CMR among metal-poor GCs of 
different GCs systems has been reported (Harris et al. 2006, Strader 
et al. 2006, Mieske et al. 2006). Despite this trend has been interpreted as a 
mass-metallicity relation, it has also been suggested that the origin of the 
blue tilt might be related to photometric errors: blue GCs might be more 
extended systems and, as a consequence, marginally resolved by ACS-HST (Kundu 
2008). Forte et al. (2007) have detected through ground-based observations, a 
notable tilt of the colours of the blue GCs associated with NGC\,4486 (M87) in 
the Virgo cluster. They found a linear relation that corresponds to a 0.06 mag 
$(C-T_1)$ colour increase per magnitude. This finding is against the hypothesis
of the blue tilt arising from photometric errors. Recently, Romanowsky et al. 
(2008) have arrived at a similar conclusion.

\section{The colour-magnitude relation in the Antlia cluster}

Smith Castelli et al. (2008) have found a tight CMR in the Antlia cluster. 
Considering new members identified through GEMINI-GMOS spectroscopy, this 
relation spans $\sim 10$ mag in brightness, from giant elliptical to dwarf 
galaxies, with no apparent change of slope (Smith Castelli 2008, Smith Castelli
et al. in preparation). This slope is consistent with those found in other 
clusters like Virgo and Fornax, and is also similar to that found by Forte et 
al. (2007) for the blue tilt of NGC\,4486 GCs. When all these stellar systems 
are shifted to the Antlia distance, they seem to follow a common CMR from blue 
GCs to giant ellipticals as it can be seen in Fig. 1 (Smith Castelli et al. in 
preparation). The equation of the Antlia CMR fit is $T_{1_0}=39.2(\pm1.6)-14.1(\pm0.9)*(C-T_1)_0$. Considering $(m-M)=32.73$ for Antlia (Dirsch et al. 2003) 
and $(m-M)=31.0$ for NGC\,4486 (Forte et al. 2007), the equation of the 'blue 
tilt' of NGC\,4486 shifted to the Antlia distance reads 
$T_{1_0}=43.2-16.7*(C-T_1)_0$.

\section{The ``blue tilt'' in other globular cluster systems}

We have found no evidence of a blue tilt in the GC system of NGC\,1399, the 
central galaxy of the Fornax cluster, but a unimodal colour distribution for 
the brightest GCs (Dirsch et al. 2003, Forte et al. 2007). Similarly, Bassino 
et al. (2008) detected no blue tilt in the blue GCs around the two dominant 
galaxies of the Antlia cluster, NGC\,3258 and NGC\,3268 (but see Harris et al. 
2006); both GC systems present unimodal colour distributions for the most 
luminous clusters. 

However, the blue tilt has been observed in some normal elliptical and S0 
galaxies (Mieske et al. 2006, Spitler et al. 2006). The fact that this feature 
is present in some galaxies and not in others might be pointing to an 
environmental effect regarding the formation of the GCs, that is not clearly
understood yet.

\section{Concluding remarks and future perspectives}

There is a uniform CMR for early-type galaxies in nearby clusters (Sandage 
1972, Bower, Lucey \& Ellis 1992, Smith Castelli et al. 2008, Misgeld et al. 
2008, De Rijcke et al. 2008). The blue tilt of the NGC\,4486 GC system appears 
to extend this relation to very faint magnitudes. However, this effect is not 
seen if we consider the GC system of NGC\,1399 in Fornax and the brightest 
galaxies in Antlia.\\

Assuming that the CMR of early-type galaxies corresponds to a mass-metallicity 
relation, it sets a limit for the highest metallicity that a member galaxy of 
a cluster can reach for a given mass. The colour (metallicity?) range covered
by early-type galaxies in Antlia is similar to the one corresponding to all 
NGC\,4486 GCs.\\

A common and well defined CMR among giant and dwarf elliptical
galaxies is surprising given that their chemical histories are
supposed to be different. Moreover it would be if such a relation
extends towards the GCs regime. We have no explanation
for such extended universal trend at the moment. On one hand, the fact that 
the 'blue tilt' is seen in some galaxies and not in others, is pointing to an 
environmental effect. On the other hand, the universality of the CMR of 
early-type galaxies lead several authors to suggest that its build up is 
independent of the local environment (e.g. Smith Castelli et al. 2008). If 
blue GCs follow this universal trend, is it a consecuence of the environment 
in which they are now inmersed or due to their primordial formation 
conditions?\\ 

We plan to further extend this investigation before arriving at sensible 
conclusions. The existence of a common CMR between early-type galaxies and 
blue GCs would represent a new challenge to our understanding of the formation 
and evolution of stellar systems in the Universe.

\acknowledgments
This work was funded with grants from CONICET, ANPCyT and 
UNLP (Argentina). TR is grateful for support from the Chilean Center for 
Astrophysics, FONDAP No. 15010003.

\begin{referencias}
\vskip 1cm
                                                                                
\reference Baum W., 1959, PASP, 71, 106
\reference Bassino L. P., Richtler T., Dirsch B., 2008, MNRAS, 386, 1145
\reference Bower R. G., Lucey J. R., Ellis R. S., 1992, MNRAS, 254, 601 
\reference De Rijcke S., Penny S., Conselice C., Valcke S., Held E. V., 2008, MNRAS, 393, 798
\reference Dirsch B., Richtler T., Geisler D., et al., 2003, AJ, 125, 1908
\reference Forte J. C., Faifer F. R., Geisler D., 2007, MNRAS, 382, 1947
\reference Harris W. E., Whitmore B. C., Karakla D. et al., 2006, ApJ, 636, 90
\reference Kundu A., 2008, AJ, 136, 1013
\reference Kuntschner H., 2000, MNRAS, 315, 184
\reference Mieske S. et al., 2006, ApJ, 653, 193 
\reference Mieske S., Hilker M., Infante L., Mendes de Oliveira C., 2007, A\&A, 463, 503
\reference Misgeld I., Mieske S., Hilker M., 2008, A\&A, 486, 697
\reference Rakos K., Schombert J., Maitzen H. M., Prugovecki S., Odell A., 2001, AJ, 121, 1974
\reference Romanowsky A. J., Strader J. Spitler L. R. et al., 2009, AJ, accepted (astro-ph/0809.2088)
\reference Sandage A., 1972, ApJ, 176, 21
\reference Smith Castelli A. V., 2008, PhD Thesis, Fac. de Cs. Astronómicas y Geofísicas, Universidad Nacional de La Plata, Argentina
\reference Smith Castelli A. V., Bassino L. P., Richtler T., Cellone S. A., Aruta C., Infante L., MNRAS, 386, 2311
\reference Spitler L. R., Larsen S. S., Strader J. et al., 2006, AJ, 132, 1593
\reference Strader J., Brodie J. P., Spitler L., Beasley M. A., 2006, AJ, 132, 2333
\reference Strader J., Smith H. G., 2008, AJ, 136, 1828
\end{referencias}

\begin{figure}
\center
\includegraphics[scale=0.6]{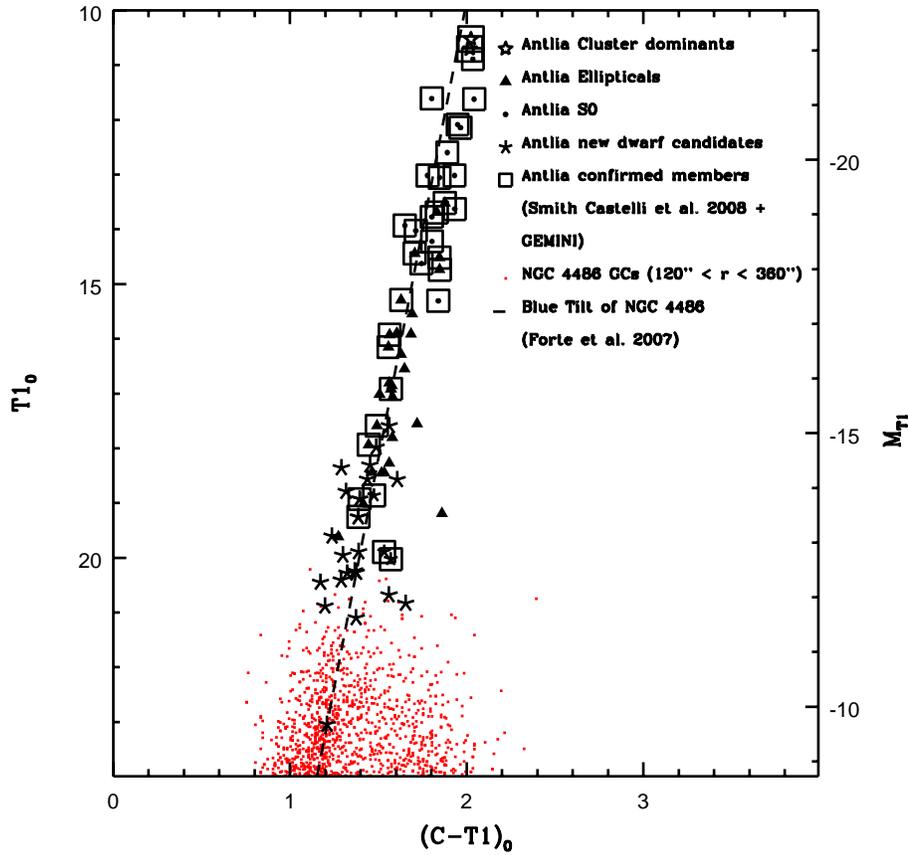}
\caption{Colour-Magnitude diagram of early-type galaxies in the Antlia cluster,
and NGC\,4486 globular clusters (red points) shifted to the Antlia distance. 
As a reference, the dashed line is the fit to the blue tilt of NGC\,4486 
globulars found by Forte et al. (2007).}
\end{figure}

\end{document}